\documentclass[12pt]{article}
\usepackage{times}
\usepackage{geometry}
\usepackage[utf8]{inputenc}
\usepackage{enumitem,amssymb}
\usepackage{ragged2e}
\usepackage{graphicx}
\usepackage{comment}
\usepackage{multicol}
\usepackage[usenames]{xcolor} 
\definecolor{xlinkcolor}{cmyk}{1,1,0,0}
\usepackage{url}
\usepackage[
 colorlinks=true,    
 linkcolor=xlinkcolor,     
 citecolor=xlinkcolor,     
 filecolor=xlinkcolor,  
 urlcolor=xlinkcolor,      
 final=true
]{hyperref}
\usepackage[super,sort&compress]{natbib}
\usepackage{enumitem}
\setenumerate{itemsep=0mm}

\usepackage{hyperxmp}
\usepackage{hyperref}
\usepackage{authblk}

\def\dbar{\overline{D}{}^0}

\def\ra{\!\rightarrow\!}
\def\cp{$CP$}
\def\CP{$CP$}

\begin{document}

\renewcommand\Affilfont{\itshape\small}

\author[1]{Y.~Amhis}\affil[1]{LAL, Universit\'{e} Paris-Sud, CNRS/IN2P3, Orsay, France}
\author[2]{Sw.~Banerjee}\affil[2]{University of Louisville, Louisville, Kentucky, USA}
\author[3]{E.~Ben-Haim}\affil[3]{LPNHE, Sorbonne Universit\'e, Paris Diderot Sorbonne Paris Cit\'e, CNRS/IN2P3, Paris, France}
\author[4]{M.~Bona}\affil[4]{School of Physics and Astronomy, Queen Mary University of London, London, UK}
\author[5]{A.~Bozek}\affil[5]{H. Niewodniczanski Institute of Nuclear Physics, Krak\'{o}w, Poland}
\author[6]{C.~Bozzi}\affil[6]{INFN Sezione di Ferrara, Ferrara, Italy}
\author[5]{J.~Brodzicka}
\author[7]{M.~Chrzaszcz} \affil[7]{European Organization for Nuclear Research (CERN), Geneva, Switzerland}
\author[8]{J.~Dingfelder}\affil[8]{University of Bonn, Bonn, Germany}
\author[9]{U.~Egede}\affil[9]{Monash University, Melbourne, Australia}
\author[10]{M.~Gersabeck}\affil[10]{School of Physics and Astronomy, University of Manchester, Manchester, UK}
\author[11]{T.~Gershon}\affil[11]{Department of Physics, University of Warwick, Coventry, UK}
\author[12]{P.~Goldenzweig}\affil[12]{Institut f\"ur Experimentelle Teilchenphysik, Karlsruher Institut f\"ur Technologie, Karlsruhe, Germany}
\author[13]{K.~Hayasaka}\affil[13]{Niigata University, Niigata, Japan}
\author[14]{H.~Hayashii}\affil[14]{Nara Women's University, Nara, Japan}
\author[7]{D.~Johnson}
\author[15]{M.~Kenzie}\affil[15]{Cavendish Laboratory, University of Cambridge, Cambridge, UK}
\author[16]{T.~Kuhr}\affil[16]{Ludwig-Maximilians-University, Munich, Germany}
\author[17]{O.~Leroy}\affil[17]{Aix Marseille Univ, CNRS/IN2P3, CPPM, Marseille, France}
\author[18,19]{H.~B.~Li}\affil[18]{Institute of High Energy Physics, Beijing 100049, People’s Republic of China}\affil[19]{University of Chinese Academy of Sciences, Beijing 100049, People’s Republic of China}
\author[20,21]{A.~Lusiani}\affil[20]{Scuola Normale Superiore, Pisa, Italy}\affil[21]{INFN Sezione di Pisa, Pisa, Italy}
\author[14]{K.~Miyabayashi}
\author[22]{P.~Naik}\affil[22]{H.H.~Wills Physics Laboratory, University of Bristol, Bristol, UK}
\author[23]{T.~Nanut}\affil[23]{Institute of Physics, Ecole Polytechnique F\'{e}d\'{e}rale de Lausanne (EPFL), Lausanne, Switzerland}
\author[9]{M.~Patel}
\author[24,25]{A.~Pompili}\affil[24]{Universita' di Bari Aldo Moro, Bari, Italy}\affil[25]{INFN Sezione di Bari, Bari, Italy}
\author[26]{M.~Rama}\affil[26]{INFN Sezione di Pisa, Pisa, Italy}
\author[26]{M.~Roney}\affil[26]{University of Victoria, Victoria, British Columbia, Canada}
\author[27]{M.~Rotondo}\affil[27]{Laboratori Nazionali dell'INFN di Frascati, Frascati, Italy}
\author[23]{O.~Schneider}
\author[28]{C.~Schwanda}\affil[28]{Institute of High Energy Physics, Vienna, Austria}
\author[29]{A.~J.~Schwartz}\affil[29]{University of Cincinnati, Cincinnati, Ohio, USA}
\author[30,31]{B.~Shwartz}\affil[30]{Budker Institute of Nuclear Physics (SB RAS), Novosibirsk, Russia}\affil[31]{Novosibirsk State University, Novosibirsk, Russia}
\author[17]{J.~Serrano}
\author[32]{A.~Soffer}\affil[32]{Tel Aviv University, Tel Aviv, Israel}
\author[33]{D.~Tonelli}\affil[33]{INFN Sezione di Trieste, Trieste, Italy}
\author[34]{P.~Urquijo}\affil[34]{School of Physics, University of Melbourne, Melbourne, Victoria, Australia}
\author[35]{R.~Van Kooten}\affil[35]{Indiana University, Bloomington, Indiana, USA}
\author[36]{J.~Yelton}\affil[36]{University of Florida, Gainesville, Florida, USA}

\title{ {\it Snowmass 2021 Letter of Interest:} \\
  Decays of Heavy Flavors Beauty, Charm, and Tau \\
  \vskip0.30in \Large The Heavy Flavor Averaging Group} 
\date{}
\normalsize
\maketitle

\vspace{-0.35in}

\noindent {\large \bf Corresponding Author:} \\
\noindent {A.~J.~Schwartz (University of Cincinnati), alan.j.schwartz@uc.edu} \\

\noindent {\large \bf Thematic Area(s):}

\noindent Rare Processes and Precision Measurement Frontier  \\
\noindent $\blacksquare$ (RF01) Weak Decays of $b$ and $c$ \\
\noindent $\blacksquare$ (RF04) Baryon \& Lepton Number Violation \\
\noindent $\blacksquare$ (RF05) Charged Lepton Flavor Violation ($e$, $\mu$, and $\tau$) 

\vspace{0.4in}

\noindent {\large \bf Abstract:}

\vspace{0.2in}

The Heavy Flavor Averaging Group provides this Letter of Interest (LOI) as
input to the Snowmass 2021 Particle Physics Community Planning Exercise organized
by the Division of Particles and Fields of the American Physical Society. 
Research in heavy flavor physics is an essential component of particle physics,
both within and beyond the Standard Model. To fully realize the potential of this field,
we advocate strong support within the U.S. high energy physics program for ongoing and
future experimental and theory research in heavy flavor physics.

\clearpage


The Heavy Flavor Averaging Group (HFLAV) is an international collaboration
responsible for calculating world averages of $b$-hadron, $c$-hadron and
$\tau$-lepton properties from relevant experimental measurements. All results are
documented in an extensive collection of web pages~\cite{HFLAV_web} and compiled
into a biannual review article. The most recent compilation will appear in
{\it Eur. J. Phys. C}.\cite{HFLAV_preprint} Many of our
world averages are used by the Particle Data Group.\cite{PDG} 
With this perspective we comment on the importance of heavy flavor physics
as input to the Snowmass 2021 Particle Physics Community Planning Exercise.

Flavor physics is one of the cornerstones of the Standard Model (SM).
Indeed, many advances in the construction of the SM resulted from
flavor physics research. This includes the three-generation prediction of the
Kobayashi-Maskawa mechanism, the universality of the gauge interactions, the high
masses of the top quark and the weak gauge bosons, and the presence of
large charge-parity (\CP) violation in beauty hadrons. Similarly,
flavor physics can give insights into physics beyond the SM, referred
to here as new physics (NP). Some specific examples are:
\begin{itemize}
\item The baryon asymmetry of the universe necessitates \cp\ violation
  well beyond that provided by the SM. Precise measurements of \cp\ violation
  in heavy flavor ($b$, $c$, and $\tau$) decays may uncover new sources of \cp\
  violation. Processes in which the SM predicts zero or very small \cp\ violation
  are particularly sensitive to NP amplitudes.

\item The origin of the three generations of fermions and their Yukawa couplings are
  not explained by the SM; this shortcoming hints that there might exist additional physics.
  These couplings are probed by precision measurements of the Cabibbo-Kobayashi-Maskawa (CKM)
  matrix, which test the three-generation picture and the prediction that quark-flavor
  non-universality depends on only three real parameters and one complex phase in the
  CKM matrix.

\item The SM gauge couplings depend only on weak isospin. This can be tested
  to high precision in weak decays of $D$ and $B$ mesons, and of $\tau$ leptons;
  any deviation would be a sign of NP. Thus, measurements of lepton-flavor
  non-universality, lepton-flavor violation, and lepton-number violation
  are important probes of NP. Similarly, flavor-changing neutral currents,
  which necessarily proceed via internal loop diagrams, are sensitive to
  the presence of new intermediate states and couplings.
\end{itemize}

The sensitivity of heavy flavor measurements leads to tight constraints
on NP, in many cases at energy scales far above those accessible at an
energy-frontier machine. Some flavor physics measurements have shown
deviations from SM predictions and in fact might indicate NP, e.g.,
measurements of the lepton-universality ratios $R(D)$, $R(D^*)$, and
$R^{}_K$. In these cases, more data and confirmation among different
experiments are needed to firmly establish these discrepancies.
If these deviations were due to NP, additional deviations
should arise in other measurements of $B^{}_{(s)}$ and $\tau$ decays.\cite{Feruglio}
Some flavor physics measurements have motivated searches for new mediators
at the LHC;\cite{Sirunyan} in this case heavy flavor physics played a role
at an energy-frontier machine.

In addition to elucidating the structure of weak interactions, heavy flavor physics
provides a unique laboratory for studying strong interactions. In particular,
beginning in 2003 a number of hadronic states containing $c$ or $b$ quarks
with unexpected quantum numbers and properties have been discovered, e.g.,
the $D^+_{sJ}$, $X(3872)$, $Z_c^+(4430)$, and $P_c^+$ (pentaquark) states.
These discoveries opened up a new research area by revealing
new ways in which QCD forms bound states.

Heavy flavor physics has been, and continues to be, a major effort in the US
high energy physics (HEP) program. Flavor physics was the main thrust of the
CLEO, Belle, and BaBar experiments, and a significant thrust of the CDF, D\O, and
SLD experiments. It was a main thrust of the Fermilab fixed target program.
US groups played a leading role in studying $B$ and $D$ physics at Belle,
and also at LEP experiments. Today, the leading flavor physics
experiments are Belle~II, LHCb, and BES-III, and the number of
US groups working on these are 18, 6, and 5, respectively.
Flavor physics is also studied at the energy-frontier
experiments CMS and ATLAS. In addition,
US theory groups play leadership roles in flavor physics
phenomenology and lattice QCD calculations. As in other areas
of HEP, data analysis and theory calculations 
are often performed by small groups at universities and laboratories
and can take years to complete. Thus, long-term support of such groups
is essential for advancing the field and training the next generation
of particle physicists.

The majority of heavy flavor physics measurements for the next ten years will be made by
Belle~II, an $e^+e^-$ experiment, and LHCb, which runs at the CERN LHC. The physics
programs of both experiments are well-documented.\cite{BII-Physics, LHCb-PII-Physics}
BES-III will also make many measurements.\cite{BESIII-Physics} 
Belle~II builds upon the extremely successful physics programs of the ``$B$-factory''
experiments Belle and BaBar, while LHCb continues its very successful physics program
achieved with LHC Runs I and II data. Belle~II excels at reconstructing
$B$, $D$, and $\tau$ decays with undectected neutrinos, photons, and
neutral mesons in the final state, while LHCb benefits from large
production cross sections for both mesons and baryons, and (due to a large Lorentz boost)
excellent decay-time resolution. The two experiments can be viewed as complementary, e.g., 
LHCb will make high-statistics measurements of $B^0_s$ decays, while Belle~II will study
$\tau$ decays. Both experiments will study $B$ and $D$ electroweak penguin decays,
one typically with high statistics (LHCb) and one typically with low backgrounds (Belle~II).
The resulting systematic uncertainties will be quite different. In addition
to complementarity, there is also notable symbiosis among flavor physics experiments.
For example, BES-III produces $D^0$-$\dbar$ pairs in a quantum-correlated state and
uses such correlations to measure strong phase differences between $D^0\ra K^{}_S\pi^+\pi^-$
and $\dbar\ra K^{}_S\pi^+\pi^-$ decays. These strong phase differences will subsequently
be used by Belle~II and LHCb to precisely measure the CKM unitarity triangle
angle~$\phi^{}_3$ (or $\gamma$).

Finally, we point out that each generation of flavor-physics experiments involves significant
advances in detector design and performance. The Belle~II detector includes a state-of-the-art
pixel detector and an innovative particle identification detector that extends
the concept of BaBar's DIRC detector by adding focusing and precision timing. The LHCb detector
for the upcoming Run III uses a new state-of-the-art vertex detector and the next generation of
triggering architecture and hardware.
Belle II has recently begun taking data and plans to collect 50 times more data than
Belle did by the end of the decade.~\cite{BII-Physics} On the same time scale,
LHCb will collect five times more data than its current sample. Beyond Run 4 of the LHC,
LHCb proposes an upgrade to collect an order of magnitude more data.\cite{LHCb_upgrade}
Similarly, Belle~II and the SuperKEKB accelerator have begun exploring how to extend the
data sample beyond the planned 50~ab$^{-1}$, and possibly polarize the electron beam to
expand the physics program further.

In summary, heavy flavor physics has played a crucial role in understanding the SM and in
searching for NP.
This will continue to be the case in the future. To fully realize the potential of heavy
flavor physics, we advocate that the US continue to play a leading role in the field, and
for US funding agencies to strongly support this physics.


\end{document}